\documentclass[11pt]{article}

\usepackage{times,epsfig,amsfonts,amsmath,amssymb}

\newtheorem{theorem}{Theorem}

\newtheorem{observation}[theorem]{Observation}

\long\def\denspar #1\densend {#1}

\newcommand{\ignore}[1]{}

\def\squarebox#1{\hbox to #1{\hfill\vbox to #1{\vfill}}}

   { \hspace*{\fill} \vrule width.2cm height.2cm depth0cm\smallskip}

\newif\ifcomment \commentfalse

\def\commentOFF{\commentfalse}
\long\outer\def\BC#1\EC{\ifcomment \sloppy \par \# \ldots\dotfill
{\tt #1} \dotfill \# \par \fi }

\commentOFF

\newcommand{\Int}{\operatorname{int}}

\newcommand{\conv}{\operatorname{conv}}
\newcommand{\cone}{\operatorname{cone}}

\newcommand{\Vol}{\operatorname{Vol}}

\def\RR{\mathbb R}
\def\ZZ{\mathbb Z}
\def\be{\mathbf e}
\def\bzero{\mathbf 0}

\newcommand{\hide}[1]{}

\date{}

\begin{document}

\bibliographystyle{plain}
\title{Upper Bound on the Number of Vertices of Polyhedra with $0,1$-Constraint Matrices. }
\author{
Khaled Elbassioni \thanks{
      Max-Planck-Institut fur Informatik
Stuhlsatzenhausweg 85 66123 Saarbr?cken Germany
     {\tt email: elbassio@mpi-sb.mpg.de }}
    \and
Zvi Lotker \thanks{ Centrum voor Wiskunde en Informatica Kruislaan
413, NL-1098 SJ Amsterdam The Netherlands.
      {\tt email: lotker@cwi.nl.}}
\and Raimund Seidel \thanks {Department of Computer Science
Universit\"{u}at des Saarlandes 66123 Saarbr\"{u}cken, Germany
      {\tt email: rseidel@cs.uni-sb.de }}
}

\maketitle
\begin{abstract}
In this note we show that the maximum number of vertices in
any polyhedron $P=\{x\in \mathbb{R}^d~:~Ax\leq b\}$ with $0,1$-constraint matrix $A$ and a real vector $b$ is at most $d!$.
\end{abstract}

\section{Introduction}
Let $A\in\{0,1\}^{m\times d}$ be a $0,1$-matrix and
$b\in\mathbb{R}^m$ be a given real vector. Consider the polyhedron
$P=P(A,b)=\{x\in \mathbb{R}^d~:~Ax\leq b\}$.
Denote by $EXT(P)$ the set of vertices of the polyhedron $P$.
The purpose of this short note is to show the following upper bound on the size of $EXT(P)$.

\begin{theorem}\label{t-main}
For any $A\in\{0,1\}^{m\times d}$ and any
$b\in\mathbb{R}^m$, the
number vertices of the polyhedron
$P=\{x\in \mathbb{R}^d~:~Ax\leq b\}$ is at most $d!$.
\end{theorem}

The bound of Theorem \ref{t-main} is tight: consider the polyhedron:
$$
P=\Pi_{d-1}+\RR_-=\conv\{\sigma\in\ZZ^d:~\mbox{$\sigma$ is permutation of $[d]$}\}+\cone\{-\be_1,\ldots,-\be_d\},
$$
i.e. the polyhedron with all permutation of $[d]$ as vertices (known as {\it Permutahedron}) and the negative of the unit vectors as extreme directions.
Then $P$ has $d!$ vertices and it is known \cite{Z} that $P$ can have the following
linear description with a $0,1$-constraint matrix $A$:
$$
P=\{x\in\RR^d~:~\sum_{i\in S}x_i\le \sum_{i=1}^{|S|}(n-i+1),~~\mbox{for all}~~ S\subseteq[d]\}.
$$

In the next section we give a few preliminaries that will be needed to prove the Theorem \ref{t-main}. In Section \ref{s3}, we give the proof of the theorem.

\section{Preliminaries}

First, we may assume without loss of generality, by translating $P=P(A,b)$ if necessary,
that the vector $b$ is strictly positive.
To deal with the fact that $P$ is unbounded, we define
$\hat{P}=P\cap\{x\in \mathbb{R}^d~:~\sum_{i=1}^d x_i\geq -M\}$, for a sufficiently big $M$.
Note that $EXT(\hat{P})\setminus EXT(P)\subseteq\{x\in\mathbb{R}^d~:~\sum_{i=1}^dx_i=-M\}$, and that $0\in\Int(P)\cap\Int(\hat{P})$ since $b>0$,
where for $X\subseteq\mathbb{R}^d$, we denote by $\Int(P)$ the interior of $P$.
We need the following simple facts, which are well-known and also are not difficult to prove.
\begin{observation}\label{l2}
Let $Q$ be a $d$-dimensional polytope and $z\in\Int(Q)$ be a point in the interior of $Q$.
Let $F_1=\conv\{p_1,\ldots,p_r\}$ and $F_2= \conv\{q_1,\ldots,q_s\}$ be two facets of $Q$, where
$p_1,\ldots,p_r,q_1,\ldots,q_s\in EXT(Q)$ are vertices of $Q$. Let $\lambda_1,\ldots,\lambda_r$ and $\mu_1,\ldots,\mu_r$ be positive real numbers. Then
\begin{equation*}\label{pyr}
\Int(\conv\{z,\lambda_1p_1,\ldots,\lambda_rp_r\})\cap\Int(\conv\{z,\mu_1q_1,\ldots,\mu_rq_s\})=\emptyset.
\end{equation*}
\end{observation}

\begin{observation}[\cite{FKR2000}]\label{l3}
The volume $\Vol^d(Q)$ of any $d$-dimensional polytope $Q$ with integer vertices is an integer multiple of
$\frac{1}{d!}$. In particular $\Vol^d(Q)\geq \frac{1}{d!}$.
\end{observation}


\bigskip

Given a polytope   $Q$ which contains $\bzero$ as an interior point, the
{\it polar} polytope   $Q^{\circ}$ is defined as
$$
Q^{\circ}=\{x\in\mathbb{R}^d~:~y^Tx\leq 1~~\forall y\in Q\},
$$
see e.g. \cite{S}. It is known that $Q^{\circ}$ also contains $\bzero$ as an interior point, and that the vertices and facets of $Q$ are in
one-to-one correspondence with the facets and vertices of $Q^{\circ}$, respectively.

\section{Proof of Theorem \ref{t-main}} \label{s3}
Consider $(\hat{P})^{\circ}$, the polar of the polytope $\hat{P}$ defined by adding a bounding inequality to the polyhedron $P=P(A,b)$ as above.
Form polarity, it follows that if $F=\{x\in\mathbb{R}^d~:~x\in P,~~a_i^Tx=b_i\}$ is a facet of $P$, where $a_i^T$ is the $i$th row of $A$ and $b_i$ is the $i$th component of $b$,
then the corresponding polar vertex $v_i$ in $EXT((\hat{P})^{\circ})$ lies on the ray connecting the origin to $a_i$, at a distance $d(a_i,\bzero)=\frac{1}{b_i}$ from the origin.

Note that $(\hat{P})^{\circ}$ contains exactly one negative vertex $u=(-\frac{1}{M},\ldots,-\frac{1}{M})$ which does not correspond to a facet of $P$. It follows then that the vertices of $P$ are in one-to-one correspondence with the facets of $(\hat{P})^{\circ}$ that do not contain $u$.
Consider any two such facets $F_1$ and $F_2$. Then $F_1=\conv\{\frac{1}{b_{i_1}}a_{i_1},\ldots,\frac{1}{b_{i_r}}a_{i_r}\}$ and
$F_2=\conv\{\frac{1}{b_{j_1}}a_{j_1},\ldots,\frac{1}{b_{j_s}}a_{j_s}\}$, where $i_1,\ldots,i_r$ and $j_1,\ldots,j_s$ are indices from $[m]$. Now, since $\bzero\in\Int(\hat{P}^{\circ})$ and $F_1$ and $F_2$ are facets of $\hat{P}^{\circ}$, we note by Observation \ref{l2} that
$$
\Int(\conv\{\bzero,a_{i_1},\ldots,a_{i_r}\})\cap\Int(\conv\{\bzero,a_{j_1},\ldots,a_{j_s}\})=\emptyset.
$$
By Observation \ref{l3}, we have that the volume $\Vol^d(S(a_{i_1},\ldots,a_{i_r})\})$ of the polytope $S(a_{i_1},\ldots,a_{i_r})=\conv\{\bzero,a_{i_1},\ldots,a_{i_r})$ arising by "stretching" the vertices of a face $F=\conv\{\frac{1}{b_{i_1}}a_{i_1},\ldots,\frac{1}{b_{i_r}}a_{i_r}\}$ of $\hat{P}^{\circ}$ is at least $1/d!$. Since all such polytopes $S(a_{i_1},\ldots,a_{i_r})$ lie completely inside the $d$-dimensional hypercube, whose volume is $1$, we conclude that there number, which is equal to the number of facets of $\hat{P}^{\circ}$ is at most $d!$. This finishes the proof.

\bigskip

It can be easily seen that the above proof can be extended to matrices $A\in\{-K,\ldots,-1,$ $0,1,\ldots,K\}^{m\times d}$, in which case the corresponding upper-bound on the number of vertices will be $(2K+1)^d d!$. In particular, if the number of bits to
represent a given integer constraint matrix is $L$, then the number of vertices of the polyhedron
described by $A$ is at most $(2^L+1)^d d!$.

\end{document}
danny_milman@hotmail.com
s} 21 (2000), 121-130.

\bibitem{S}

Alexander Schrijver, {\it Theory of Linear and Integer Programming}, John Wiley and Sons, 1986.

\bibitem{Z}

Gunter M. Ziegler, {\it Lectures on Polytopes}, Graduate Texts in Mathematics, No 152 (Springer, Berlin, 1995).

\end{thebibliography}

\end{document}

danny_milman@hotmail.com